\documentclass{article}
\usepackage[margin=0.55in]{geometry}
\makeatletter
\def\thanks#1{\protected@xdef\@thanks{\@thanks
\protect\footnotetext{#1}}}
\makeatother
\date{}
\usepackage[dvipsnames]{xcolor}

\usepackage{cite}
\usepackage{amsmath,amssymb,amsfonts}
\usepackage{graphicx}
\usepackage{textcomp}
\usepackage{algorithm}
\usepackage{algpseudocode}
\usepackage{cancel}

\makeatletter
\def\BState{\State\hskip-\ALG@thistlm}
\makeatother

\newcommand{\mua}{\mu_{{\mathrm a}}}  
\newcommand{\bmua}{\bar{\mu}_{{\mathrm a}}}
\newcommand{\hmua}{\hat{\mu}_{\rm a}}
\newcommand{\hbmua}{\hat{\bar{\mu_{{\mathrm a}}}}}

\newcommand{\mus}{\mu'_{{\rm s}}}
\newcommand{\bmus}{\bar{\mu}_{\rm s}'}
\newcommand{\hmus}{\hat{\mu}'_{\rm s}}
\newcommand{\hbmus}{\hat{\bar{\mu_{{\mathrm s}}}}'}

\newcommand{\coefunit}{{\rm mm}^{-1}}

\newcommand{\dontshow}[1]{}

\newif\iflt
\ltfalse 
\definecolor{rd}{rgb}{0,0,0} 
\definecolor{rd2}{rgb}{0,0,0} 

\newcommand{\mm}{ {\rm mm}}
\newcommand{\mussup}[1]{\mu^{'{#1}}_{{\rm s}}}

\begin{document}
\title{A model-based iterative learning approach for diffuse optical tomography}
\author{Meghdoot Mozumder, Andreas Hauptmann, 
Ilkka Nissilä, Simon R. Arridge, and Tanja Tarvainen
\thanks{
This work was supported in part by The Finnish Cultural Foundation (project 00200746), Academy of Finland (projects (314411, 312342, 336799, 336796, 320166, 334817, \textcolor{rd}{338408}), and the CMIC-EPSRC platform grant (EP/M020533/1). {\em (Corresponding author: Meghdoot Mozumder).}}
\thanks{Meghdoot Mozumder is with the
Department of Applied Physics, University of Eastern Finland, P.O. Box 1627, 70211 Kuopio, Finland (e-mail: meghdoot.mozumder@uef.fi). }
\thanks{Andreas Hauptmann is with the Research Unit of Mathematical Sciences; University of Oulu, Oulu, Finland and also with the Department of Computer Science, University College London, London WC1E 6BT, United Kingdom.}
\thanks{Ilkka Nissilä is with the Department of Neuroscience and Biomedical Engineering, Aalto University, Aalto University School of Science, P.O. Box 12200, FI-00076 Aalto, Espoo, Finland.}
\thanks{Simon R. Arridge is with the Department of Computer Science, University College London, London WC1E 6BT, United Kingdom.}
\thanks{Tanja Tarvainen is with the Department of Applied Physics, University of Eastern Finland, P.O. Box 1627, 70211 Kuopio, Finland and also with the Department of Computer Science, University College London, London WC1E 6BT, United Kingdom.}}

\maketitle

\begin{abstract}
Diffuse optical tomography (DOT) utilises near-infrared light for imaging spatially distributed optical parameters, typically the absorption and scattering coefficients. The image reconstruction problem of DOT is an ill-posed inverse problem, due to the non-linear light propagation in tissues and limited boundary measurements. The ill-posedness means that the image reconstruction is sensitive to measurement and modelling errors. The Bayesian approach for the inverse problem of DOT offers the possibility of incorporating prior information about the unknowns, rendering the problem less ill-posed. It also allows marginalisation of modelling errors utilising the so-called Bayesian approximation error method. A more recent trend in image reconstruction techniques is the use of deep learning \iflt\cancel{techniques}\fi, which has shown promising results in various applications from image processing to tomographic reconstructions. In this work, we study the non-linear DOT inverse problem of estimating the \textcolor{rd}{(absolute)} absorption and scattering coefficients utilising a `model-based' learning approach, essentially intertwining learned components with the model equations of DOT. The proposed approach was validated with 2D simulations and 3D experimental data. We demonstrated improved absorption and scattering estimates for targets with a mix of smooth and sharp image features, implying that the proposed approach could learn image features that are difficult to model using standard Gaussian priors. Furthermore, it was shown that the approach can be utilised in compensating for modelling errors due to coarse discretisation enabling computationally efficient solutions. Overall, the approach provided improved computation times compared to a standard Gauss-Newton iteration.
\end{abstract}


\section{Introduction}\label{sec:introduction}

Diffuse optical tomography (DOT) utilises boundary measurements of near-infrared light to estimate spatially distributed optical absorption and scattering parameters in biological tissues \cite{arridge1999optical,gibson2005recent,durduran2015}. The distribution of these optical parameters is useful in obtaining information on tissue function and structure with applications, for example, in imaging of breast cancer \cite{grosenick2016,cochran2019hybrid}, \textcolor{rd}{prostate imaging \cite{he2019clinical,jiang2011trans}, } 
neonatal brain imaging \cite{hebden2002three}, functional imaging of the adult brain \cite{hoshi2016,wheelock2019high}, and pre-clinical small animal imaging \cite{chen2012optical}. 

The image reconstruction problem of DOT, is an ill-posed inverse problem. The ill-posedness means that even small errors in measurements or modelling can cause large errors in the image reconstruction. An established strategy to handle \iflt\cancel{the} \fi the ill-posedness of DOT image reconstruction has been to use regularisation techniques. These techniques utilised assumptions such as smoothness of the solution \cite{arridge1993performance}\textcolor{rd}{, sparsity \cite{shaw2014,bhowmik2016dimensionality},} or its derivative (total-variation) \cite{paulsen1996enhanced} to obtain stable inversion. 
In a similar manner, Bayesian estimation utilises prior probability distributions of the unknowns, based on previously available knowledge, to compute the posterior probability distribution as a solution to the inverse problem \textcolor{rd}{ \cite{ye1999optical,Kaipio,diamond2006}.} 
In this regard, the Bayesian approximation error (BAE) approach has become a standard computational technique in ill-posed inverse problems such as DOT \cite{Kaipio,Arridge2006}. \textcolor{rd}{The BAE approach computes statistics of modelling errors, such as reduced model or uncertainties, to compensate these during the solution of the inverse problem, for example \cite{Heino2005,Arridge2006,Tarvainen2010,Mozumder2014}.
For more information on image reconstruction problem of DOT and various methodologies, see e.g. \cite{arridge2009,durduran2015,hoshi2016,shaw2014,diamond2006} and the references therein. }



Recently, deep learning methods have shifted the focus of tomographic imaging from classical, purely model-based techniques to data-driven approaches. A significant influence of these techniques has been the availability of large measurement databases, advances in computing capabilities and the potential to learn image features from the data itself. These have \iflt\cancel{lead} \fi \textcolor{rd}{led} to major improvements for many linear inverse problems, where high quality reference reconstructions can be readily obtained. Most notably, these include X-ray computed tomography \cite{jin2017deep,kang2017deep,adler2017solving}, magnetic resonance imaging \cite{schlemper2017deep,hammernik2018learning,hauptmann2019real} and photoacoustic tomography \cite{hauptmann2018model,davoudi2019deep,hauptmann2020deep}. 

For non-linear inverse problems and in particular in DOT, deep learning methods are scarce. This can be explained in parts by the difficulty to obtain high quality training data, costly model evaluations, and limitations of direct reconstruction approaches. Therefore, the initial applications of deep learning methods for DOT have been based on directly learning a non-linear mapping from the boundary measurement data to the spatially distributed optical coefficients \cite{feng2018back,fan2019solving,sabir2020convolutional} where the reconstruction operator was learned either by \iflt\cancel{a classic}\fi \textcolor{rd}{classical} neural networks \cite{feng2018back} \iflt\cancel{by} \fi or using convolutional neural networks (CNNs) 
\cite{fan2019solving,sabir2020convolutional}. 
The approach was utilised in estimating absolute absorption coefficients \cite{feng2018back}, difference scattering coefficients \cite{fan2019solving}, and spatially-constant values of both the coefficients \cite{sabir2020convolutional}.  
The methods were validated with simulations \cite{feng2018back,fan2019solving} and with two homogeneous experimental phantoms  \cite{sabir2020convolutional}. Improved estimates in terms of estimation accuracy and computation time were reported.

In contrast to directly learning a reconstruction operator, one can combine deep learning with model-based approaches. This can offer the possibilities to overcome some inherent limitations of the learning-based approaches, such as biases due to training samples and need for large training datasets \cite{hauptmann2018model}. 
\textcolor{rd}{Furthermore, model-based approaches are improved by providing complementary prior information and information on model uncertainties by training neural-networks to learn image features from a database of target images}. 
A straightforward possibility to include the model equations into the reconstruction is to utilise CNNs as a post-processing tool after an initial reconstruction is obtained, for instance by removing streaking artefacts from filtered back-projection in computerised tomography \cite{kang2017deep, jin2017deep} or sharpening D-bar reconstructions in electrical impedance tomography \cite{hamilton2018deep,hamilton2019beltrami}. Alternatively, an architecture to invert the Lippmann-Schwinger equation for difference imaging of absorption coefficients in DOT was developed in \cite{yoo2019deep}. 

In this work, we follow the approach of iterative model-based techniques \cite{adler2017solving,hammernik2018learning,hauptmann2018model} where learned components, given for instance by a CNN, are intertwined with the model equation. In particular, we extend the iterative model-based approach in \cite{hauptmann2018model} 
for solving an inverse problem with a non-linear forward operator.  
This enables us to tackle the absolute imaging problem of reconstructing both absorption and scattering coefficients in DOT. The image reconstruction problem is solved with a Gauss-Newton algorithm augmented with deep learning. 
To our knowledge, this is the first study of simultaneous reconstruction of absolute absorption and scattering parameters in DOT utilising deep learning. The use of the iterative model-based learning was chosen because of 1) non-linearity of the inverse problem, 2) to emphasise generalisability of the inversion method, and 3) to learn non-trivial features of images that are difficult to account for using conventional regularisation methods or Gaussian priors. 
\textcolor{rd}{In this approach, model correction is performed implicitly by the network while computing the iterative updates. As such, model-based estimates are enhanced by learned CNN components to obtain more accurate estimates. 
Further, learning the step length selection parameter, required for updating the estimates in each iteration, provides an accelerated iteration.} 

The rest of the paper is organised as follows. An introduction to DOT, Bayesian approach to inverse problems, and the proposed model-based learning approach is presented in Sec. \ref{sec:methods}. Implementation of the proposed approach is presented in Sec. \ref{sec:implementation}. The numerical simulations and experiments are described in Secs. \ref{sec:simu}  and \ref{sec:expt}. These are followed by discussion in Sec. \ref{sec:discussion} and conclusions in Sec. \ref{sec:conclusions}. 

\section{Methods}\label{sec:methods}

\subsection{Diffuse optical tomography}

In a typical DOT measurement setup, near-infrared light is introduced into an object from its boundary. Let $\Omega \subset \mathbb{R}^d, \, (d=2 \, \rm{or} \, 3)$ denote the domain with boundary $\partial \Omega$ where $d$ is the (spatial) dimension of the domain. In a diffusive medium, like soft biological tissue, the commonly used light transport model for DOT is the diffusion approximation to the radiative transfer equation \cite{Ishimaru}. Here, we consider the frequency-domain version of the diffusion approximation \cite{arridge1999optical}
\begin{equation}\label{deeqn}
 \left(-\nabla \cdot \frac{1}{{d}(\mua(r)+\mus(r))} \nabla  + \mua(r) + \frac{{\rm j}\omega}{c} \right) \Phi(r) = 0, \hspace{0.1mm} r \in \Omega,
\end{equation}
\begin{equation}\label{deeqnbn}
\Phi(r)+\frac{1}{2\zeta}\frac{1}{d(\mua (r)+\mus(r))} \alpha \frac{\partial \Phi(r)}{\partial \hat{n}} = \left\{ \begin{array}{ll}
         \frac{q}{\zeta}, & r \in s\\
         0, & r \in \partial \Omega \setminus s \end{array} \right. ,
\end{equation}
where $\Phi(r)$ is the photon fluence, $\mua(r)$ is the absorption coefficient, $\mus(r)$ is the (reduced) scattering coefficient, j is the imaginary unit, $\omega$ is the angular modulation frequency of the input signal and $c$ is the speed of light in the medium. The parameter $q$ is the strength of the light source at location $s \subset \partial\Omega$, operating at angular modulation frequency $\omega$. Further, the parameter $\zeta$ is a dimension-dependent constant ($\zeta$ = $1/\pi$ when $\Omega \subset \mathbb{R}^{2}$, $\zeta$ = $1/2$ when $\Omega \subset \mathbb{R}^{3}$) and $\alpha$ is a parameter governing the internal reflection at the boundary $\partial \Omega$, and $\hat{n}$ is an outward \textcolor{rd}{unit vector} normal to the boundary. 
The measurable data on the boundary of the object,  exitance $\Gamma(r)$, is given by
\begin{equation}\label{bdd}
\Gamma(r) = -\frac{1}{d(\mua (r)+\mus(r))} \frac{\partial \Phi(r)}{\partial \hat{n}} = \frac{2\zeta}{\alpha}\Phi(r). 
\end{equation}

The numerical approximation of the forward model (\ref{deeqn})-(\ref{bdd}) is typically based on a finite element (FE) approximation \cite{arridge1999optical}. In the FE-approximation, the domain $\Omega$ is divided into N$_e$ non-overlapping elements joined at N$_n$ vertex nodes. The photon fluence in the finite dimensional basis is given by
\begin{equation}\label{phidisc}
\Phi^h = \sum\limits_{k=1}^{{\rm N}_n}\phi_k\psi_k(r)
\end{equation}
where $\psi_k$ are the nodal basis functions of the FE-mesh and $\phi_k$ is photon fluence in the nodes of the FE-mesh. We write the finite dimensional approximations for $\mua(r)$ and $\mus(r)$ as
\begin{align}
\label{eq:femappx_mua}
\mua(r) & \approx \mua^{h}(r) =  \sum\limits_{l=1}^{{\rm N}_{\rm n}}\mu_{{\rm a},l}\psi_l(r) \\
\label{eq:femappx_mus}
\mus (r) & \approx \mussup{h}(r) =  \sum\limits_{l=1}^{{\rm N}_{\rm n}}\iflt\cancel{\mu_{{\rm s},l}}\fi\textcolor{rd}{\mu'_{{\rm s},l}}\psi_l(r)
\end{align}
where $\mu_{{\rm a},l}$,\iflt\cancel{$\mu_{{\rm s},l}$}\fi\textcolor{rd}{$\mu'_{{\rm s},l}$} denote the absorption and scattering at the nodes of the FE-discretisation. 

Typical data types for frequency-domain DOT are the logarithm of amplitude and phase\textcolor{rd}{, which is obtained from the real and imaginary parts of the logarithm of complex exitance $\Gamma = \mathcal{A}\exp({\rm i}\phi)$, as}
\begin{equation}\label{meas}
y = \left(\begin{array}{c}
{\rm Re} \log(\Gamma)\\
{\rm Im} \log(\Gamma)
\end{array} \right) \textcolor{rd}{= \left(\begin{array}{c}
\log(\mathcal{A})\\
\phi
\end{array} \right), }
\end{equation}
where $y \in \mathbb{R}^{\rm N_m}$ is the data vector\textcolor{rd}{, $\mathcal{A}$ is the amplitude and $\phi$ is the phase delay of the measured signal}. The FE-approximation of (\ref{deeqn})--(\ref{bdd}) and (\ref{meas}) is denoted by $A_h$ and the observation model is written as
\begin{equation}\label{fwdmodel}
y = A_h(\mua,\mus)\:+\:e
\end{equation}
where $e \in \mathbb{R}^{\rm N_m}$ models the random noise in measurements, $\mua = [ \mu_{\rm{a},1},\ldots,\mu_{\rm{a},\rm{N_{\rm{n}}}} ] \in \mathbb{R}^{\rm N_n}$ and $\mus = [\iflt\cancel{\mu_{\rm{s},1},\ldots,\mu_{\rm{s},\rm{N_{\rm{n}}}}}\fi\textcolor{rd}{\mu'_{\rm{s},1},\ldots,\mu'_{\rm{s},\rm{N_{\rm{n}}}}} ] \in \mathbb{R}^{\rm N_n}$ are discretised absorption and scattering coefficients. The sub index $h$ in the mapping $A_h$ is a mesh parameter controlling the level of discretization. The operator $A_h(\mua,\mus)$ converges to the continuous forward operator as $h \to 0\; \textrm{and}\; \textrm{N}_n \to \infty $.

\subsection{Bayesian estimation}

In the Bayesian approach to inverse problems, all the parameters are considered as random variables and the uncertainties of their values are encoded into probability density models \cite{Kaipio}. Let us consider the observation model  (\ref{fwdmodel}). The solution of the inverse problem is the posterior probability density which is obtained through Bayes' theorem, and can be written as
\begin{equation}
    \pi(\mua,\mus|y) \propto \pi(y|\mua,\mus)\pi(\mua,\mus)
\end{equation}
where $\pi(y|\mua,\mus)$ is the likelihood density and $\pi(\mua,\mus)$ is the prior density. 

Since we aim at computationally efficient solutions, we compute point estimate(s) from the posterior density, the most typical choice being the {\it maximum a posteriori} (MAP) estimate. 
Assuming that the unknowns $\mua$ and $\mus$ and noise $e$ are mutually independent and Gaussian distributed, i.e. 
\begin{equation}\label{eq:gauss}
\mua \sim \mathcal{N}(\eta_{\mu_{\rm a}},\Gamma_{\mu_{\rm a}}), \,\, \mus \sim \mathcal{N}(\eta_{\mu'_{\rm s}},\Gamma_{\mu'_{\rm s}}), \,\, e \sim \mathcal{N}(\eta_{e},\Gamma_{e}), \nonumber 
\end{equation}
where $\eta_{\mu_{\rm a}}, \eta_{\mu'_{\rm s}}$ and $\eta_{e}$ are the means, and $\Gamma_{\mu_{\rm a}},\Gamma_{\mu'_{\rm s}}$ and $\Gamma_{e}$ are the covariance matrices, the MAP estimate is obtained as
\begin{equation}
\begin{split}
    \label{eq:map}
 (\hat{\mu}_{\rm a},\hat{\mu}'_{\rm s}) & =  \operatorname*{\arg \, \min}_{
 {\mua,\mus}} \left\lbrace \left\Vert L_{e}(y-A_h(\mua,\mus) - \eta_{e}) \right\Vert^2 \right. \\ & \left. \quad + \left\Vert L_{\mu_{\rm a}}(\mua-\eta_{\mu_{\rm a}}) \right\Vert^2 + \left\Vert L_{\mu'_{\rm s}}(\mus-\eta_{\mu'_{\rm s}})\right\Vert^2 \right\rbrace
\end{split}
\end{equation}
where  $L_{\mu_{\rm a}}^{\rm T}L_{\mu_{\rm a}}=\Gamma_{\mu_{\rm a}}^{-1}, L_{\mu'_{\rm s}}^{\rm T}L_{\mu'_{\rm s}}=\Gamma_{\mu'_{\rm s}}^{-1}$ and $L_{e}^{\rm T}L_{e}=\Gamma_{e}^{-1}$. The mimisatisation problem can be solved using an iterative method, such as Gauss-Newton method with iteration of the form
\begin{equation}\label{eq:gn_update}
(\hat{\mu}_{\rm a},\hat{\mu}'_{\rm s})_{i+1} = (\hat{\mu}_{\rm a},\hat{\mu}'_{\rm s})_{i} + s_i(\delta\hat{\mu}_{\rm a},\delta\hat{\mu}'_{\rm s})_{i}
\end{equation}
where $(\hat{\delta\mu}_{\rm a},\hat{\delta\mu}'_{\rm s})_{i}$ is given by
\begin{equation}\label{eq:deri}
\begin{split}
    \begin{pmatrix} \delta\hat{\mu}_{\rm a} \\ \delta\hat{\mu}'_{\rm s}
\end{pmatrix}_i
    & =     \left( J_i^{\rm T}\Gamma_e^{-1}J_i^{\rm T} + \begin{pmatrix} \Gamma_{\mu_{\rm a}}^{-1} & 0 \\ 0 &  \Gamma_{\mu'_{\rm s}}^{-1}
\end{pmatrix}\right)^{-1} \\
& \left(J_i^{\rm T}\Gamma_e^{-1}\left(y-A_h(\hat{\mu}_{{\rm a},i},\hat{\mu}'_{{\rm s},i})-\eta_e \right) \right. \\
& \left. + \Gamma_{\mu_{\rm a}}^{-1} \left(\hat{\mu}_{{\rm a},i}-\eta_{\mu_{\rm a}}\right) + \Gamma_{\mu'_{\rm s}}^{-1}\left(\hat{\mu}'_{{\rm s},i}-\eta_{\mu'_{\rm s}}\right) \right)
\end{split}
\end{equation}
and  $s_i$ is the step length parameter. Here Jacobian $J_i$ is the discrete representation of the Fr{\'e}chet derivative of the non-linear mapping $A_h(\hat{\mu}_{{\rm a},i},\hat{\mu}'_{{\rm s},i})$ at the current iterate. 

\subsection{Prior}
In this work, we use the following two Gaussian forms as our prior distributions for \iflt\cancel{asborption} \fi \textcolor{rd}{absorption} and scattering.

\subsubsection{Gaussian Ornstein-Uhlenbeck prior}
To model smooth parameter distributions, we chose the prior model for the unknown parameters (\ref{eq:gauss}) as the Ornstein-Uhlenbeck process, which belongs to the Mat{\'e}rn class of covariance functions \cite{rasmussen2006}. Ornstein-Uhlenbeck prior is a Gaussian distribution with the covariance matrix $\Gamma$ defined as 
\begin{equation}\label{prior}
    \Gamma_{\mu,mk} = \sigma_{\mu}^2\exp\Big(-\frac{\|r_m-r_k\|}{\ell}\Big)
\end{equation}
where $\mu$ denotes the unknown parameters (absorption and scattering), $\sigma_{\mu}$ is the standard deviation, $r_m$ and $r_k$ are the locations of the FE-discretisation nodes $m$ and $k$, and $\ell$ is the characteristic length scale which controls the spatial range of correlation. The prior supports correlation between neighborhood discretization points, promoting distributions that can be locally close to homogeneous.

In this work, the means of the prior were $\eta_{\mua}=  0.01 \, \coefunit$ and $\eta_{\mus}= 1 \, \coefunit$, and the standard deviations were $\sigma_{\mua} =  0.0033 \, \coefunit$ and  \iflt\cancel{$\sigma_{\mua}$} \fi \textcolor{rd}{$\sigma_{\mus}$} $ =  0.33 \, \coefunit$. The correlation length was $\ell =  8 \, {\rm mm}$.

\subsubsection{Gaussian sample-based prior} For targets which consisted of both non-smooth and smooth features, Gaussian sample-based priors were constructed using a set of sample images $\{ \mu_{\rm a},\mu'_{\rm s}\}^j, j = 1,\hdots,N_{\rm samp}$. The prior means and covariances were computed as
\begin{align}
    \eta_\mu & = \frac{1}{N_{\rm samp}}\sum_{i=1}^{N_{\rm samp}}\mu^j \\ \Gamma_\mu & = \frac{1}{N_{\rm samp}-1}\sum_{i=1}^{N_{\rm samp}}(\mu^j-\eta_\mu)(\mu^j-\eta_\mu)^{\rm T}.
\end{align}

\subsection{Bayesian approximation error method} \label{sec:bae}
In practical applications, the use of a sufficiently dense discretization may be infeasible due to computational resource and time limitations. In such a case, the observation model (\ref{fwdmodel}) with a forward operator with a fine discretisation is replaced by an approximate model 
\begin{equation}\label{approx}
y = A_H(\bmua,\bmus)+e 
\end{equation}
where the discretization parameter $H > h$ and $(\bmua,\bmus)$ are the corresponding discretised optical coefficients with discretisation $H$. In the Bayesian approximation error approach, instead of just using the approximate model (\ref{approx}) we re-write the observation model (\ref{fwdmodel}) in the following way 
\begin{equation}\label{eq:aem}
y = A_H(\bmua,\bmus) + \underbrace{\{ A_h(\mua,\mus)-A_H(\bmua,\bmus) \}}_{\varepsilon} + e.
\end{equation}
Here, $\varepsilon$ is the discretisation error describing the discrepancy between the accurate forward model and the approximate model. The Bayesian approximation error method carries out an approximate marginalisation of the posterior over the error $\varepsilon$. Following \cite{Kaipio,Arridge2006}, the MAP estimate with the Bayesian approximation error model is obtained as
\begin{equation}\label{eq:mapaem}
\begin{split}
(\hbmua,\hbmus) & =  \operatorname*{\arg \, \min}_{
{\mua,\mus}} \left\lbrace  \left\Vert L_{\varepsilon+e}(y-A_H(\bmua,\bmus)-\eta_{\varepsilon}-\eta_e) \right\Vert^2 \right. \\
& \left. \quad +  \left\Vert L_{\mu_{\rm a}}(\bmua-\eta_{\mu_{\rm a}}) \right\Vert^2 +  \left\Vert L_{\mu'_{\rm s}}(\bmus-\eta_{\mu'_{\rm s}})\right\Vert^2 \right\rbrace
\end{split}
\end{equation}
where $\eta_{\varepsilon}$ and $\Gamma_{\varepsilon}$ are the mean and covariance of the approximation error. Further,  $L_{\varepsilon+e}^{\rm T}L_{\varepsilon+e}$ = $(\Gamma_\varepsilon+\Gamma_e)^{-1}$. 
In the following sections, we refer to the solution of (\ref{eq:mapaem}) 
as the MAP estimate with the Bayesian approximation error (BAE) approach.

\subsection{Model-based learning using deep Gauss-Newton}

\begin{figure*}[tb!]
\centering
\includegraphics[scale=0.55,trim={1mm 0mm 3mm 0mm},clip]{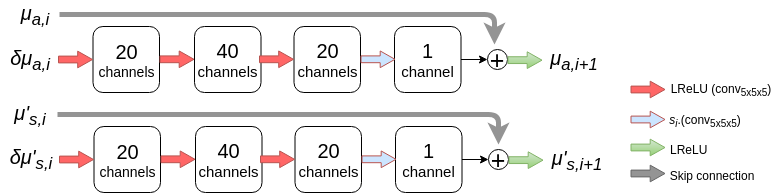}
\caption{Diagram of one CNN denoted as $G_{\theta_i}$, representing one iteration of the deep Gauss-Newton. The red arrows denote a convolutional layer with 5$\times$5 kernel for a 2D image (or 5$\times$5$\times$5 kernel for a 3D image), bias and followed by a LReLU. The resulting channels in each layer are indicated in the squares. The blue arrow denotes a convolutional layer followed by a scalar multiplication. The residual update (by the skip connection) is then projected to the positive numbers by the last LReLU (shown as green arrow). \label{fig:network}}
\end{figure*}

In this work, instead of the regular Gauss-Newton update, Eq. (\ref{eq:gn_update}), we propose to learn an update
function for each iteration
\begin{equation}\label{eq:dgn_update}
(\hat{\mu}_{\rm a},\hat{\mu}'_{\rm s})_{i+1} = G_{\theta_i}((\hat{\mu}_{\rm a},\hat{\mu}'_{\rm s})_{i}, (\delta\hat{\mu}_{\rm a},\delta\hat{\mu}'_{\rm s})_{i}).
\end{equation}
The functions $G_{\theta_i}$ correspond to CNNs with different, learned parameters ${\theta_i}$ but with the same architecture.
This implies that the update of the optical parameters including the step length selection is now learned from the data during training.
The network structure is kept simple and shown in Fig. \ref{fig:network}. 
Due to the representation of each update by a CNN applied to the optical parameters $(\hmua,\hmus)_i$ the Gauss-Newton search directions $(\delta\hat{\mua},\delta\hat{\mus})_i$, we refer to the
algorithm as a deep Gauss-Newton (DGN) \iflt\cancel{, similarly as in Ref. \cite{hauptmann2018model}}\fi. In contrast to the previously proposed model-based learning technique for DOT difference imaging in Ref. \cite{yoo2019deep}, we train the DGN layer by layer (layer corresponding here to one iterate), due to the non-linear nature of the absolute imaging problem. Hence, we learn the parameters $\theta_i$ for each iteration separately.

\section{Implementation}\label{sec:implementation}

The Toast++ software \cite{schweiger2014toast++} was utilised in the FE-solution of the diffusion equation using MATLAB (R2017b, Mathworks, Natick, MA). 
A Python library, Tensorflow (version 1.14.0) \cite{abadi2016tensorflow} was utilised in implementation and training of the DGN algorithm. 
The simulations, were carried out in a Fujitsu Celcius W550 desktop workstation, with Intel\textregistered Xeon(R) W-2125 CPU @ 4.00GHz$\times$8. 
The training of the DGN were carried out on a NVIDIA Tesla V100 GPU.

\subsection{Implementation of the deep Gauss-Newton}

The architecture chosen for the CNNs performing the update in Eq. (\ref{eq:dgn_update}) is illustrated in Fig. \ref{fig:network}. In each iteration,  optical coefficients and the corresponding search directions are given as an input to a pipeline. 
They are expanded to 20 and then 40 channels by a convolutional layer with kernel size of 5 pixels, and dimension $d=2 \, \rm{or} \, 3$, including bias and equipped with a \textcolor{rd}{`leaky'} rectified linear unit \textcolor{rd}{(LReLU)} as non-linearity, that was defined as
\iflt\cancel{$
{\rm ReLU}(\mu) = {\rm max}(\mu, 0).
$\\}\fi
$$
\textcolor{rd}{{\rm LReLU}(\mu) = {\rm max}(\mu, 0.1\mu).}
$$
\textcolor{rd}{As opposed to commonly used ReLU, the use of LReLU allow for negative values of input parameters $(\delta\hat{\mu}_{\rm a},\delta\hat{\mu}'_{\rm s})$. The number of channels represent the number of images that are output after each convolution and non-linear operations, representing the number of image variations to be learned. The expansive part of the network serves as a feature extractor (encoder) and the contracting part feature fusion (decoder). The kernel size is the pixel size on which the convolution filters are applied, and relates to the size of spatial variations to be learned.} The outputs of the pipelines are added together, and first reduced to 20 channels, equipped with a \iflt\cancel{ReLU}\fi \textcolor{rd}{LReLU}, and then to 1 channel without a non-linearity, but multiplication to a scalar value (representing the step length parameter $s_i$). 
The outcome is  added to the current iterate and projected to the positive numbers by a \iflt\cancel{ReLU }\fi \textcolor{rd}{LReLU}. 
Before applying the network, the optical parameters ($\mua,\mus$) were scaled \iflt\cancel{($\mua\mapsto10^3\times\mua,\mus\mapsto10\times\mus$)}\fi \textcolor{rd}{($\mua\mapsto10^2\times\mua,\mus\mapsto\mus$)} to present values in the same numerical range.

Since the main contribution of this work is not the specific neural network architecture, we used a simple architecture following Ref. \cite{hauptmann2018model}. As shown, the network structure is kept rather small with the idea that each $G_{\theta_i}$ primarily learns how to combine the current iterate parameters and the search directions, in contrast to a large post-processing network.

\subsection{Training the deep Gauss-Newton}\label{sec:training}

The network was trained by simulations utilising absorption and scattering distributions, i.e. 'ground-truth images', drawn from a prior $\{ \mu_{\rm a,true},\mu'_{\rm s,true}\}^j, j = 1,\hdots,N_{\rm samp}$ and corresponding simulated measurement data. 
The data was simulated using a FE-approximation of the diffusion approximation (\ref{deeqn})-(\ref{bdd}) and it was corrupted with additive noise. 
\textcolor{rd}{Since CNNs operate on uniform pixel domains, the images drawn in the mesh basis were interpolated to the image basis for the application of the CNN and back for simulation of (\ref{deeqn})-(\ref{bdd}). Alternatively, one can use graph structures to formulate the problem on the FE mesh directly \cite{herzberg2021graph}.}

The parameters $\theta_i$ for $i = 1,\hdots, i_{\rm max}$, were trained sequentially, where $i_{\rm max}$ was the maximum number of iterations. The parameters were trained on each iteration $i$ by minimising the `L$_2$-loss' function
\begin{equation}\label{eq:cost}
\min_{\theta_i} \sum_{j} \|\mu^j_{{\rm a},i+1}-\mu^j_{\rm a,true}\|+ \|\mu'^j_{{\rm s},i+1}-\mu'^j_{\rm s,true}\|,
\end{equation}
where $\mu^j_{i+1}$ at an iterate was given by Eq. (\ref{eq:dgn_update}). 
For the iteration, the initial guess, $\mu^j_1$ was chosen as the mean of prior. 
Thereafter, $\theta_1$ was trained to minimise the difference between $\mu^j_{\rm true}$ and $\mu^j_2$ (computed using Eq. (\ref{eq:dgn_update})) for all indices $j$. 
The procedure was repeated to train all $\theta_i$'s. 
In this work, $i_{\rm max}$ was chosen as 5, which has been a typical number of iterations required for convergence of Gauss-Newton iterations in DOT absolute imaging according to our experience.
Training of the $\theta_i$'s was carried out by minimising (\ref{eq:cost}) with the TensorFlow’s implementation of the Adam optimiser \cite{kingma2014adam} using batches of size 2, \textcolor{rd}{maximum of} 10 epochs and step size of $5\cdot10^{-4}$ (learning rate). \textcolor{rd}{The estimate $\mu^j_{i+1}$ at an iterate approaches the accurate solution when trained to a sufficiently low loss. The} \textcolor{rd}{training was terminated when the change of average L$_2$-loss (\ref{eq:cost}) was less than 0.1\% between two consecutive epochs.}

\textcolor{rd}{The training procedure is aimed to improve the conventional update of optical parameters (\ref{eq:deri}) based on image features learned from the training samples. }
\textcolor{rd}{ 
These image features and the step length selection are encoded in the learned parameters $\theta_i$.
The training procedure is summarised in Algorithm \ref{algo}.}


\begin{algorithm}
\caption{Training the deep Gauss-Newton \label{algo}}
\begin{algorithmic}[1]
\State Draw set $\{ \mu_{\rm a,true},\mu'_{\rm s,true}\}^j, j = 1,\hdots,N_{\rm samp}$, from prior.
\State Generate noisy measurement data using Eq. (\ref{fwdmodel}).
\State Set $(\hat{\mu}_{\rm a},\hat{\mu}'_{\rm s})^j_{1}$ as mean of prior (\ref{eq:gauss}), for $N_{\rm samp}$ cases.
\Function{Iterate}{}
\State $i \gets 1$
\While{$i<i_{\rm max}$}
\State Compute $(\delta\hat{\mu}_{\rm a},\delta\hat{\mu}'_{\rm s})^j_{i}$, using Eq. (\ref{eq:deri}) for $N_{\rm samp}$ \newline \hspace*{0.95cm} cases.
\Function{Train}{$(\hat{\mu}_{\rm a},\hat{\mu}'_{\rm s})^j_{i},(\delta\hat{\mu}_{\rm a},\delta\hat{\mu}'_{\rm s})^j_{i},( 
\mu_{\rm a,true},\newline \hspace*{1.1cm} \mu'_{\rm s,true})^j$}
\State Train $\theta_i$'s by minimizing Eq. (\ref{eq:cost}) 
\EndFunction{ Return $\theta_i$}
\State $(\hat{\mu}_{\rm a},\hat{\mu}'_{\rm s})_{i+1} \gets G_{\theta_i}((\hat{\mu}_{\rm a},\hat{\mu}'_{\rm s})_{i}, (\delta\hat{\mu}_{\rm a},\delta\hat{\mu}'_{\rm s})_{i})
$.
\State $i\gets i+1$
\EndWhile
\EndFunction
\end{algorithmic}
\end{algorithm}

\subsection{Evaluating the deep Gauss-Newton}

After training the parameter sets $\theta_i$, the learned iterative reconstruction scheme was evaluated by applying the network $G_{\theta_i}$ at each iteration. This procedure was equivalent to Algorithm 1, starting by setting $(\hat{\mu}_{\rm a},\hat{\mu}'_{\rm s})_{1}$, calling function `ITERATE', and skipping the function `TRAIN'.


\section{Simulations}\label{sec:simu}


\subsection{Data generation}

\begin{figure}[tb!]
\centering
\includegraphics[scale=1.2,trim={7mm 1mm 5mm 5mm},clip]{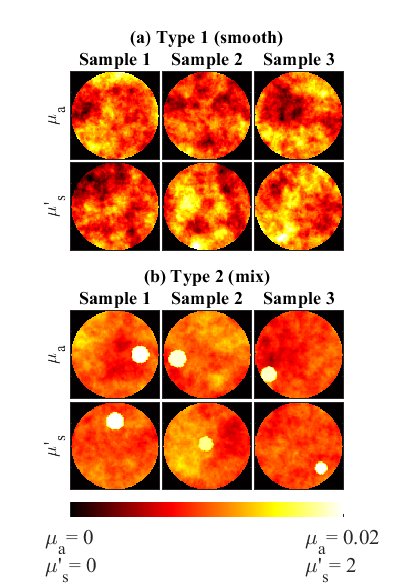}
\caption{(a) Three samples of 'smooth' absorption $\mu_{\rm a}$ and scattering $\mu_{\rm s}$ images drawn from Ornstein-Uhlenbeck prior and (b) three samples drawn from 'mix' targets. \label{fig:sample}}
\end{figure}

In the numerical studies, the domain $\Omega \subset \mathbb{R}^2$ was a circle with a radius of \iflt\cancel{$25\:\mm$}\fi\textcolor{rd}{$35 \, {\rm mm}$}. The measurement setup consisted of $16$ sources and $16$ detectors modelled as \iflt\cancel{$1\:\mm$}\fi \textcolor{rd}{$2\:\mm$} wide surface patches located at equi-spaced angular intervals on the boundary.  
The target optical parameters were either drawn from the Gaussian Orstein-Uhlenbeck prior, as shown in Fig. \ref{fig:sample} (a), or drawn as a mix of smoothly varying background (drawn from the  Orstein-Uhlenbeck prior) and sharp circular inclusions with varying contrast and radii, as shown in Fig. \ref{fig:sample} (b). 
\textcolor{rd}{The optical parameter values were chosen to mimic those in biological tissues \cite{jacques2013optical}.}

\begin{figure}[tb!]
\centering
\includegraphics[scale=0.65,trim={23mm 9mm 15mm 15mm},clip]{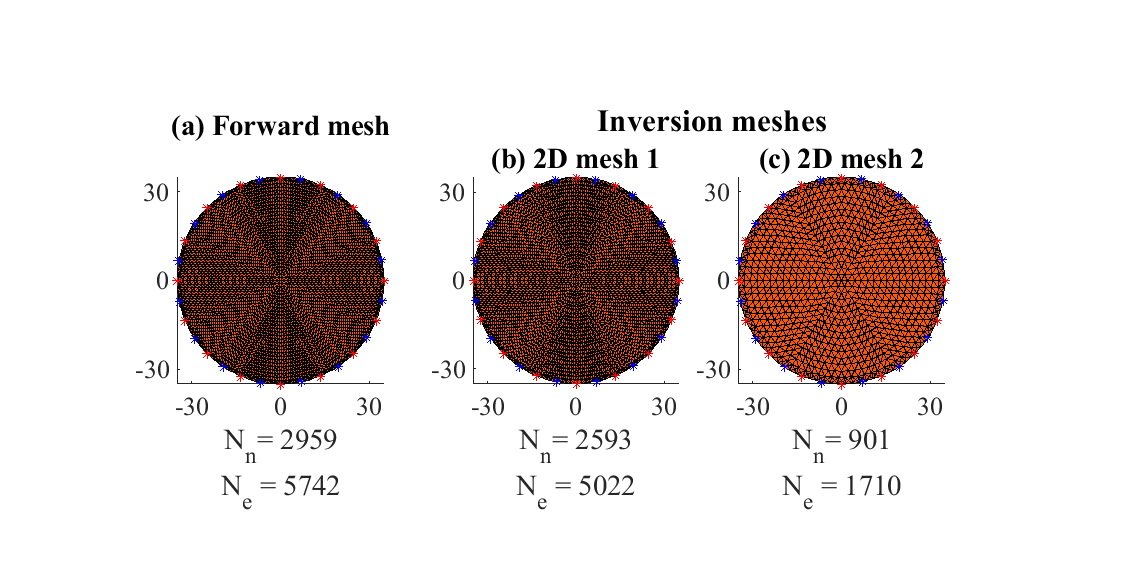}
\caption{Meshes used in 2D simulations. Locations of the sources and detectors are shown as red and blue stars on the mesh boundaries. The number of FE-nodes $N_{\rm n}$ and elements $N_{\rm e}$ for the meshes are also displayed. (a) Forward mesh was used for simulating measurement data. 
\textcolor{rd}{(b)-(c) Inversion meshes 2D mesh 1 and 2D mesh 2
were used for estimating the optical coefficients from the simulated data.} \label{fig:2dmesh}}
\end{figure}

The measurement data were simulated using the FE-approximation of the diffusion approximation (\ref{deeqn})-(\ref{bdd}), using a 'forward mesh' shown in Fig. \ref{fig:2dmesh} (a).  Random measurement noise $e$ drawn from a zero-mean Gaussian distribution 
\begin{equation}\label{eq:sim_noise}
\pi (e) = \mathcal{N}(0, \Gamma_e), \quad \Gamma_e = {\rm diag}(\sigma_{e,1}^2,\ldots,\sigma_{e,{\rm N_m}}^2),    
\end{equation} where the standard deviations $\sigma_{e,k}$ were specified as $1\%$ of the simulated noise-free data, was added to the simulated data.

\subsection{Estimation}

Estimation of the optical parameters was first carried out in a `2D mesh 1', shown in Fig. \ref{fig:2dmesh} (b). `Forward mesh' and `2D mesh 1' had a similar level of discretisation but they were chosen differently to avoid making an inverse crime \cite{Kaipio}. 

The deep Gauss-Newton (DGN) were trained with a set of 1000 absorption and scattering images as described in Section \ref{sec:training}. The training images were either smooth distributions or mix distributions or both of these. The training times are given in Table \ref{table}. 

\begin{table}[!tb]
\renewcommand{\arraystretch}{1.3}
\caption{Computation times for training 1000 `smooth' samples in hours (h) or seconds (s), using different inversion meshes. Average evaluation time per sample is reported. }
\label{table}
\centering
\begin{tabular}{|c|c|c|c|c|c|}
\hline
Inverse mesh & \multicolumn{2}{|c|}{Training} & \multicolumn{3}{|c|}{Evaluation}\\
& \multicolumn{1}{c}{DGN} & \multicolumn{1}{c|}{BAE} &  \multicolumn{1}{c}{GN} & \multicolumn{1}{c}{BAE} & \multicolumn{1}{c|}{DGN}\\
\hline
2D mesh 1 & \iflt\cancel{102h}\fi \textcolor{rd}{17h} & & 9s & & 7s \\
2D mesh 2 &  \iflt\cancel{101h}\fi  \textcolor{rd}{19h} & 127s & 4s & 4s & 3s \\
\hline
3D mesh 1 &  & & 615s & & \\
3D mesh 2 & \iflt\cancel{145h}\fi\textcolor{rd}{20h} & 3319s & 112s & 132s & 96s  \\
\hline
\end{tabular}
\end{table}

Absorption and scattering images were estimated using the model-based DGN as described in Sec. \ref{sec:implementation}. 
For comparison, reconstructions using the conventional Gauss-Newton (GN) Eq. (\ref{eq:gn_update}) were computed. 
The conventional GN method used the Orstein-Uhlenbeck prior when estimating smooth targets and the Gaussian sample-based prior when estimating mix targets. 
The DGN utilised the Gaussian Orstein-Uhlenbeck prior for all cases. 
The mean $\eta_{e}=0$ and covariance $\Gamma_e$ of the measurement noise were assumed known.

\subsubsection{Case 1: DGN trained with smooth images}
\begin{figure}[tb!]
\centering
\textbf{DGN trained with \underline{smooth} images}\\
\includegraphics[scale=1,trim={0mm 0mm 0mm 0mm},clip]{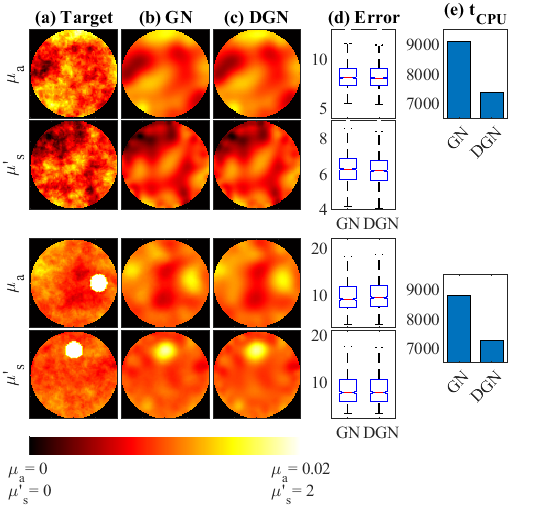}
\caption{Reconstructions of (top) smooth and (bottom) mix targets, with conventional GN and DGN trained using smooth images. (a) Target absorption $\mua$ and scattering $\mus$, (b) estimates using GN and (c) estimates using DGN. Statistics of estimation errors for 1000 evaluation cases are shown as `boxplots' in (c), and the evaluation times in (d). \label{fig:2drecon_smooth}}
\end{figure}

In this case, the proposed DGN was trained using the smooth images.
Example reconstructions from one smooth target and one mix target are shown in Fig. \ref{fig:2drecon_smooth} (a)-(c). 
Further, statistics of relative errors of absorption and scattering estimates from 1000 simulated targets and the computation times are shown in Fig. \ref{fig:2drecon_smooth} (d)-(e) and in Table \ref{table}. 

As shown, DGN provide estimates with similar quality and accuracy as the conventional GN for both smooth and mix targets. The computation times of the estimates with DGN were lower than with the GN. Although the DGN required additional evaluation of CNNs, it still provided a computational advantage since the step-length parameter ($s_i$) was learned and didn't need to be computed at each iteration.

\subsubsection{Case 2: DGN trained with mix targets}

\begin{figure}[tb!]
\centering
\textbf{DGN trained with \underline{mix} images}\\
\includegraphics[scale=1,trim={0mm 0mm 0mm 0mm},clip]{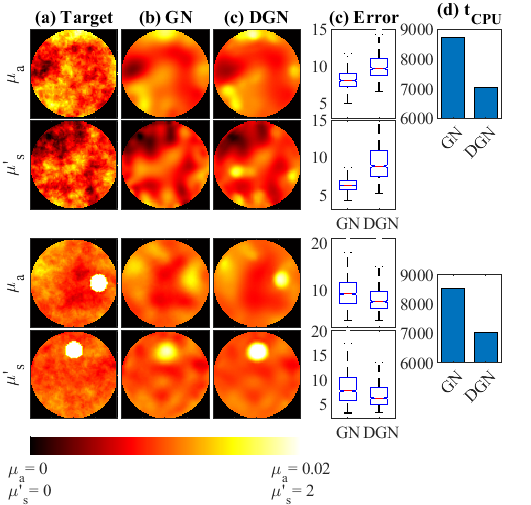}
\caption{Reconstructions of (top) mix and (bottom) smooth targets, with conventional GN and DGN trained using mix images. (a) Target absorption $\mua$ and scattering $\mus$, (b) estimates using GN and (c) estimates using DGN. Statistics of estimation errors for 1000 evaluation cases are shown in (c), and the evaluation times in (d).  \label{fig:2drecon_mix}}
\end{figure}

In this case, the DGN was trained with mix images. Example reconstructions from one smooth target and one a mix target are shown in Fig. \ref{fig:2drecon_mix} (a)-(c). 
Further, statistics of relative errors of absorption and scattering estimates from 1000 simulated targets and the computation times are shown in Fig. \ref{fig:2drecon_mix} (d)-(e) and in Table \ref{table}. 
 
As it can be seen, the absorption and scattering inclusions of the mix targets are visualised more clearly when estimated using the proposed DGN method. Furthermore, the estimates from the mix targets obtained using DGN  have lower errors when compared to the estimates obtained conventional GN. On the other hand, the `out-of-distribution' smooth targets now show higher errors, due to artificial sharpening of the estimated images with the DGN. Again, the computation times of the estimates with the DGN were lower.

\subsubsection{Case 3: DGN trained with both smooth and mix targets} In this case, the DGN was trained with both smooth and mix images, i.e. 50 \% of the training samples were smooth images and 50 \% were mix images. 

\begin{figure}[tb!]
\centering
\textbf{DGN trained with \underline{both smooth and mix} images}\\
\includegraphics[scale=1,trim={0mm 0mm 0mm 0mm},clip]{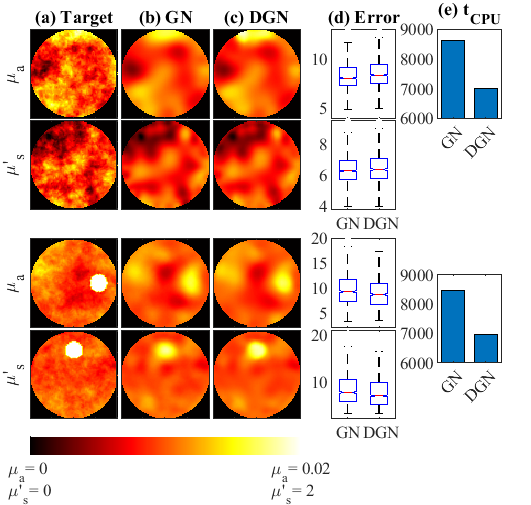}
\caption{Reconstructions of (top) smooth and (bottom) mix targets, with conventional GN and DGN trained using both smooth and mix images. (a) Target absorption $\mua$ and scattering $\mus$, (b) estimates using GN and (c) estimates using DGN. Statistics of estimation errors for 1000 evaluation cases are shown in (c), and the evaluation times in (d).  \label{fig:2drecon_both}}
\end{figure}

Example reconstructions from one smooth target and one mix target are shown in Fig. \ref{fig:2drecon_both} (a)-(c). 
Further, statistics of the relative errors of absorption and scattering estimates from 1000 simulated targets and the computation times are shown in Fig. \ref{fig:2drecon_both} (d)-(e) and in Table \ref{table}. 

As it can be seen, the reconstructions from smooth targets are similar in quality when comparing DGN and GN solutions. Also the relative errors are similar in magnitude. For the mix targets, the reconstructions obtained using DGN show slightly better contrast than those obtained with GN. However, the difference between these is not as clear as when compared to results obtained using mix training data. Further, the relative errors are similar in magnitude. Computation times of the estimates with the DGN were lower.

\subsection{Estimation in presence of 
modelling errors}

Then, the DGN method in the presence of modelling errors due to a coarse discretisation was studied. 
Use of coarse discretisation is beneficial for memory consumption and computation times.
In these simulations, a mesh, `2D mesh 2' shown in Fig. \ref{fig:2dmesh} (c), was used.

Absorption and scattering images were estimated using the DGN. 
For comparison, estimates using conventional GN and Gauss-Newton augmented with Bayesian approximation error modelling (BAE) described in Sec. \ref{sec:bae} were computed.

For training the DGN, the network parameters were trained using 1000 smooth distributions or mix distributions. For training the BAE method, the statistics of the discretisation errors were calculated with 1000 samples of smooth or mix distributions.
These absorption and scattering parameters $(\mua,\mus)$ were projected from `2D mesh 1' to `2D mesh 2' to obtain $(\bmua,\bmus)$. Thereafter, the accurate forward solutions $A_h(\mua,\mus)$ and approximate \iflt\cancel{froward} \fi \textcolor{rd}{forward} solutions $A_H(\bmua,\bmus)$ were computed. Finally, the mean ($\eta_{\varepsilon}$) and covariance ($\Gamma_\varepsilon$) of the approximation error $\varepsilon = A_h(\mua,\mus)-A_H(\bar{\mua},\bar{\mus})$ were computed.
The training times of the BAE and DGN are given in Table \ref{table}. As seen, the DGN training times were higher than training of the conventional BAE.

\subsubsection{Case 1: DGN and BAE trained with smooth images} 
\begin{figure}[tb!]
\centering
\textbf{DGN and BAE trained with \underline{smooth} images}\\
\includegraphics[scale=1,trim={2mm 4mm 0mm 2mm},clip]{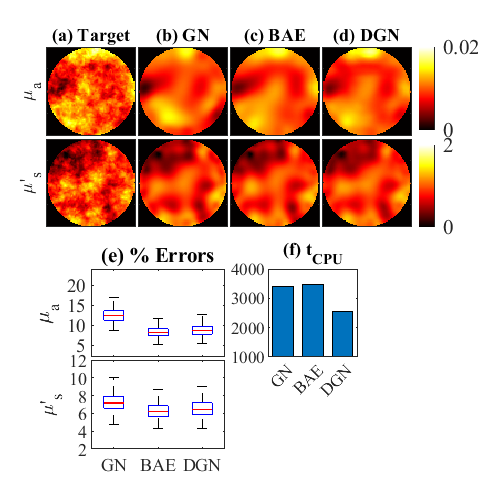}
\caption{Reconstructions of smooth targets in the presence of modelling erros. (a) Target absorption $\mua$ and scattering $\mus$,  (b) estimates using GN, (c) BAE, and (d) DGN. Artefacts due to modelling errors are seen in the GN estimates. These are compensated in the BAE and DGN estimates. Statistics of the estimation errors for 1000 evaluation cases are shown in (e), and the evaluation times in (f).  \label{fig:werr_smooth}}
\end{figure}

Example reconstructions from a smooth target in the presence of discretisation errors is shown in \ref{fig:werr_smooth} (a)-(d).
The estimates were computed using the proposed DGN and compared against conventional GN and GN \iflt\cancel{agumented} \fi \textcolor{rd}{augmented} with BAE. Further, statistics of the relative errors of the estimates and computation times from 1000 simulated targets, are  shown in Fig. \ref{fig:werr_smooth} (e)-(f) and Table \ref{table}. 

It can be seen that the proposed DGN and the BAE effectively marginalise modelling errors, providing comparable images that have lower errors when compared to the GN. The relative errors obtained with the BAE were comparable to using the DGN. As seen, the evaluation time of the DGN was lower than using the BAE, although the training time of the DGN was higher than the BAE.

\subsubsection{Case 2: DGN and BAE trained with mix images} 
\begin{figure}[tb!]
\centering
\textbf{DGN and BAE trained with \underline{mix} images}\\
\includegraphics[scale=1,trim={2mm 2mm 0mm 2mm},clip]{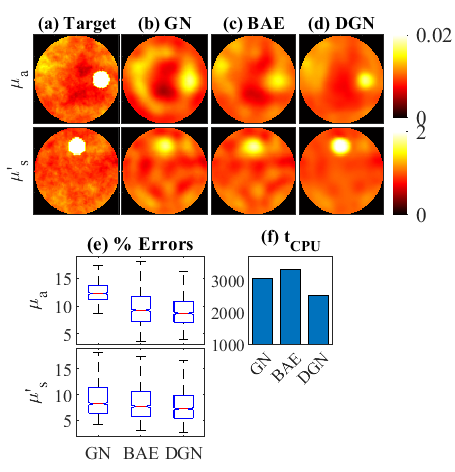}
\caption{Reconstructions of mix targets in the presence of modelling errors.(a) Target absorption $\mua$ and scattering $\mus$, (b) estimates using GN, (c) BAE, and (d) DGN. Artefacts due to modelling errors are seen in the GN estimates. These are compensated in the BAE and DGN estimates. Statistics of the estimation errors for 1000 evaluation cases are shown in (e), and the evaluation times in (f).  \label{fig:werr_mix}}
\end{figure}

Example reconstructions from a smooth target obtained using DGN, GN and BAE in the presence of discretisation errors are shown in Fig. \ref{fig:werr_mix} (a)-(d).
Statistics of the relative errors of the estimates and computation times from 1000 simulated targets are shown in Fig. \ref{fig:werr_mix} (e)-(f) and Table \ref{table}.

As it can be seen, the DGN and the BAE provide images with better quality than the conventional GN that suffers from artefacts due to the modelling errors. 
Furthermore, the relative errors of DGN and BAE are lower than of GN, with DGN providing the lowest relative errors for both absorption and scattering estimates. The results demonstrate that the DGN can compensate for modelling errors slightly better than the BAE in the case of non-smooth targets. The DGN also had lower computation time compared to BAE. However, the training time was higher.


\section{Experiments}\label{sec:expt}

\begin{figure}[tb!]
\centering
\includegraphics[scale=0.54,trim={10mm 2mm 15mm 8mm},clip]{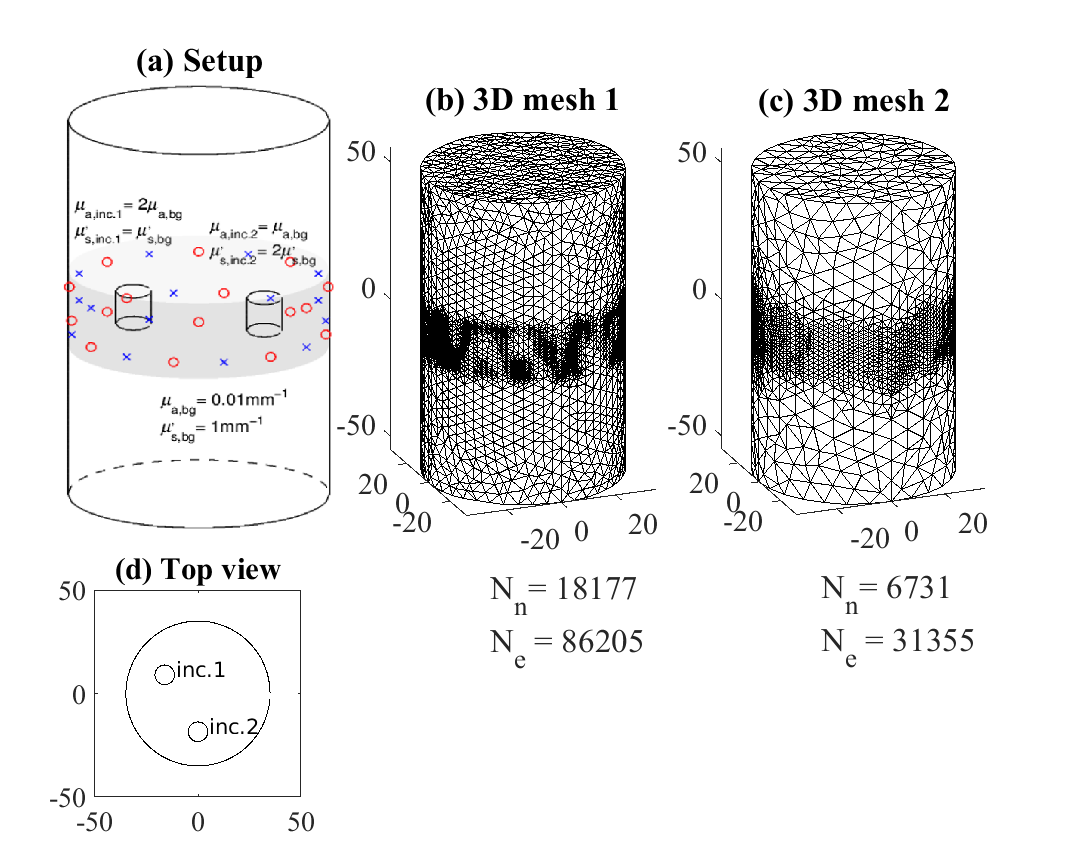}
\caption{(a) DOT experimental setup with position of sources (red circles) and detectors (blue crosses) and two cylindrical inclusions. (b) '3D mesh 1' is a densely discretised 3D mesh used in calculating 'reference' estimate with the GN method. (c) '3D mesh 2' was used in calculating estimates with the DGN, GN and GN augmented BAE. The number of FE-nodes (N$_{\rm n}$) and elements (N$_{\rm e}$) in the meshes are also displayed. \textcolor{rd}{(d) Top view of the phantom showing location of the inclusions.} \label{fig:3Dmeshes}}
\end{figure}

The phantom experiment was carried out with the frequency-domain DOT instrument at the Aalto University, Finland \cite{Nissila2005}. 
A  cylindrical phantom with a radius of 35 mm and height of 110 mm illustrated in Fig. \ref{fig:3Dmeshes} (a) \textcolor{rd}{and (d)} was studied. The background optical parameters were approximately $\mua=0.01\: \coefunit$ and $\mus=1\:\coefunit$ at wavelength 800 nm, and two cylindrical inclusions which both had the diameter and height of $9.5\:\rm{mm}$, were located such that the central planes of the inclusions coincided with the central xy-plane of the cylinder domain. 
The optical properties of the inclusion 1 were approximately $\mu_{\rm a,inc.1} = 0.02\: \coefunit$, $\mu_{\rm s,inc.1}' = 1\: \coefunit$ (i.e., purely absorption contrast) and the optical properties of the inclusion 2 were $\mu_{\rm a,inc.2} = 0.01\:\coefunit$, $\mu_{\rm s,inc.2}' = 2\:\coefunit$ (i.e., purely scatter contrast), respectively. 

\textcolor{rd}{The phantom provides absorption and scattering contrast of 2:1, similar to optical parameter variations in tumors \cite{jacques2013optical}.} The source and detector configuration in the experiment consisted of 16 sources and 15 detectors arranged in an interleaved order on two rings located 6 mm above and below the central xy-plane of the cylinder domain. The locations of sources and detectors are shown with red circles and blue crosses respectively in Fig. \ref{fig:3Dmeshes} (a). The measurements were carried out at 785 nm with an optical power of $8\:\rm{mW}$ and angular modulation frequency $\omega =100 \, {\rm MHz}$. 
The nearest measurement data from each source position were removed from measured amplitude and phase data.
Logarithm of  amplitude and phase shift were stored as data, leading to real-valued measurement vectors $y\in\mathbb{R}^{360}$.

\subsubsection{Reference estimate}

The absorption and scattering parameters were estimated in a densely discretised mesh to provide a reference for the other approaches.   
A `3D mesh 1' shown in Fig. \ref{fig:3Dmeshes} (b) was used. 

To calibrate the source strength and phase coupling of the forward model to the experimental setup, a global calibration was carried out on the measured data as the initial step in the reconstruction process. Following the initial estimation procedure in \cite{Kolehmainen2009}, the logarithm of source strength and phase coupling were modelled by additive constants. The initialisation step consisted of a four-parameter fit of global background parameters $\mu_{a_1*}$ and $\mu_{s_1*}'$ as well as a global additive shift $\eta$ of log(amplitude) data and a global additive shift $\phi$ of phase data, 
\begin{equation}\label{eq:4param}
 (\mu_{a*},\mu_{s*}',\eta,\phi) = \operatorname*{\arg \, \min}_{\mu_{a*},\mu_{s*}',\eta,\phi} \|L_e((y + \Delta y)-A(\mu_{a*},\mu_{s*}'))\|^2,    
\end{equation}
where $\Delta y = (\eta,\phi)^{\rm T}$. The initialization problem was solved using GN method which resulted in parameter estimates: $ \eta=3.5,\;\phi=0.025,\;\mu_{a_1*} = 0.01 {\rm mm}^{-1}, \mu_{s_1*}' = 0.8 {\rm mm}^{-1}.$ 

Once the initialisation was completed, the measurement data was transformed for the standard GN estimation (\ref{eq:gn_update}) by the recovered global offsets as $y \mapsto y + \Delta y$, and the initial parameter values and the prior means were set to the estimated values $\mu_{a_1*}$ and $\mu_{s_1*}'$. 

Thereafter, the reference estimates were calculated with the standard GN method.
In the solution, the measurement noise was assumed to be 1$\%$ of the measured absolute values of the log amplitude and phase.
Further, the Ornstein-Uhlenbeck prior was used with the same parameter values that were used in the simulations used.

\subsubsection{Estimates in presence of model errors}

A coarse `3D mesh 2', shown in Fig. \ref{fig:3Dmeshes} (c) was used in evaluating the proposed DGN approach in the presence of discretisation errors.
For comparison, the BAE approach was utilised, and the related minimisation problem was solved using GN.
Before reconstructions, the above described global calibration was repeated for the data using the `3D mesh 2'. 

To train the DGN, 1000 samples of 3D distributed absorption and scattering distributions were drawn from the Ornstein-Uhlenbeck prior and simulated measurement data was calculated using `3D mesh 1'. To mimic the experimental situation of unknown source strengths and phase coupling, the corresponding coefficients ($\eta,\phi$) were drawn from uniform distributions 
$$
\eta\sim\mathcal{U}(1,4), \quad \phi\sim\mathcal{U}(0.01,0.04)  
$$
and added to the simulated measurement data. Thereafter, for the training procedure, the `3D mesh 2' was used to compute the four-parameter fit using Eq. (\ref{eq:4param}), transform the data ($y \mapsto y + \Delta y$), and subsequently to train the DGN with the procedure described in Section \ref{sec:training}. 

\subsubsection{Results} 
\begin{figure}[tb!]
\centering
\includegraphics[scale=0.55,trim={3mm 7mm 3mm 3mm},clip]{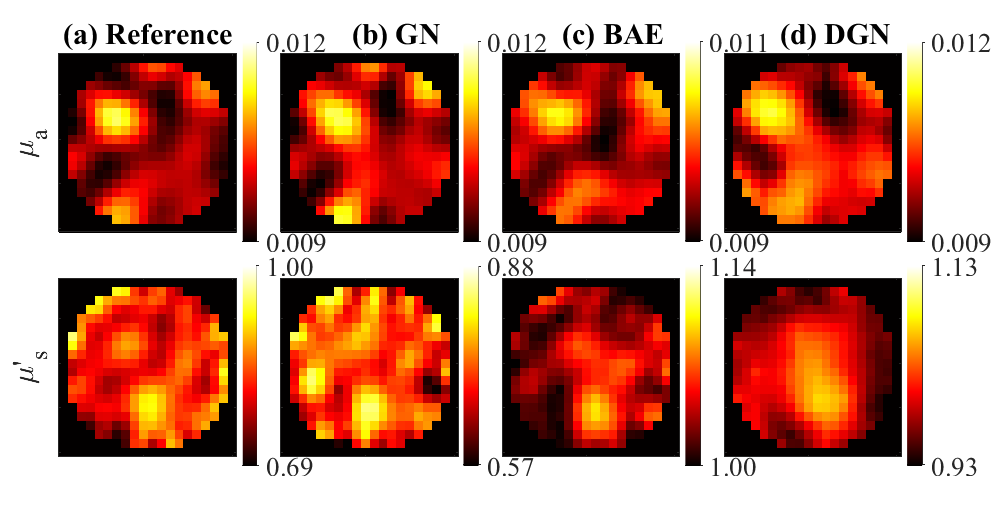}
\caption{Absorption $\mua$ and scattering $\mus$ distributions reconstructed from experimental data. (a) Reference estimate computed with dense mesh `3D mesh 1', (b) GN estimate computed with mesh `3D mesh 2' and (c) DGN estimate compute with `3D mesh 2'.
\label{fig:expt_recon}}
\end{figure}

Estimated absorption and scattering distributions calculated using the DGN, GN and GN augmented with BAE in a coarse discretisation and a GN in a fine discretisation are shown in Fig. \ref{fig:expt_recon}. 
Estimates on the central xy-plane of the cylindrical domain are visualised.

As it can be seen in Fig. \ref{fig:expt_recon} (a), reference estimates computed in `3D mesh 2' show locations of absorption and scattering inclusions with some boundary artefacts. 
Absolute and difference imaging reconstructions using the phantom were earlier presented in Refs.~\cite{Kolehmainen2009,mozumder2015nonlinear}, and they show similar quality reconstructions.
The boundary  artefacts are larger in both absorption and scattering estimates obtained with the GN in `3D mesh 2' in Fig. \ref{fig:expt_recon} (b). 
Furthermore, these artefacts are reduced utilising both the GN augmented with BAE and DGN, as seen in Fig. \ref{fig:expt_recon} (c)-(d).
 
Note that the training of the DGN utilised smooth 3D images drawn from Ornstein-Uhlenbeck prior, and not sharp inclusions as were present in the experimental phantom. As such, the evaluation in this case was carried out using a `out-of-distribution' target. The experiment demonstrates that DGN trained on generalised smooth images and coarse meshes can produce images with comparable quality to more dense meshes, using lower computational resources than BAE method.

The training and reconstruction time for the DGN, GN and the BAE method are presented in Table \ref{table}. It can be seen that the reconstruction time using the experimental data was lower using the proposed DGN, although the training time (using simulated data) was higher.

\section{Discussion}\label{sec:discussion}

This work proposed a `model-based' deep-learning approach to absolute imaging of DOT. We demonstrated that model-based learning can provide computational advantages to the standard Bayesian inversion methods, when convolutional neural networks trained with similar images are utilised in the iterative parameter estimation procedure. This was demonstrated utilising smooth and mix targets in Figs. \ref{fig:2drecon_smooth}, \ref{fig:2drecon_mix} and \ref{fig:2drecon_both}. 

In addition to the improvement in computation time, we demonstrated that the method can lead to improved estimates for targets which contain both sharp and smooth features, in Fig. \ref{fig:2drecon_mix}. As such, the proposed method can learn non-Gaussian image features more efficiently than standard Gaussian priors. We note that using hyper-priors in Bayesian estimation also allow estimating such sharp image discontinuities \cite{calvetti2008hypermodels}. However, utilising hyper-priors in DOT might lead to additional computational burden, for further estimating hyper-parameters related to the priors.

In Figs. \ref{fig:werr_smooth}, \ref{fig:werr_mix} we demonstrated the efficacy of the proposed method in compensation of discretisation errors. As shown in \ref{fig:werr_smooth}, the proposed DGN can marginalise discretisation errors with similar accuracy to BAE for smooth targets, providing an advantage in computational time. For mix targets, the proposed DGN outperforms the BAE in estimation accuracy and computational times. This is possibly because CNNs can learn and compensate non-Gaussian modelling errors more efficiently than BAE, which assumes modelling errors as Gaussian. The authors refer to \cite{lunz2020learned,smyl2021learning} for more discussions on modelling error corrections using CNNs.

The proposed approach was applied to experimental data and the estimates are presented in Fig. \ref{fig:expt_recon}. As shown, both BAE and DGN trained on smooth images can effectively compensate modelling errors, and provide accuracy comparable to reference estimates. 

The purpose of the study was to present the applicability of the approach to absolute imaging problem of DOT with one relevant architecture.
Therefore, we have not presented an in-depth assessment of different deep learning architectures in this work. The proposed DGN approach could be further improved by carrying out a systematic optimisation of the network architectures and involved parameters. Further, the different estimation scenarios, such as the discretisation level and geometry affect on the optimal training procedure as well as modelling errors and experimental system related uncertainties.
These topics are part of planned future studies.
The authors refer to \cite{schlemper2017deep,hauptmann2020deep} for other architectures that could also be relevant for DOT. 

\begin{figure}[tbp!]
\centering
\includegraphics[scale=1.2,trim={5mm 1mm 5mm 5mm},clip]{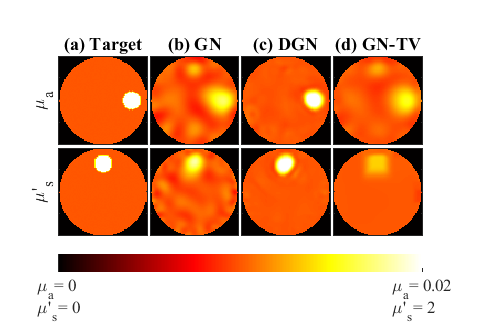}
\caption{(a) Target absorption (top) and scattering (bottom) distributions. The reconstructed distributions using (b) Gauss-Newton (GN) algorithm, (c) deep Gauss-Newton (DGN) algorithm \textcolor{rd2}{and (d) Gauss-Newton algoritm using a total-variation prior (GN-TV).} \label{fig:cross}}
\end{figure}

\textcolor{rd}{A classical challenge in DOT is the cross-talk between the optical parameters in the reconstructions. Artifacts due to cross-talk are difficult to spot in smooth or mix images, as were used in the article. 
Figure \ref{fig:cross} shows reconstructions with sharp inclusions solved with GN and DGN. The DGN was trained with sharp inclusions, with random inclusion locations, sizes and contrast. As seen, the cross-talk artefacts that are visible in GN reconstruction are not seen in DGN. }\textcolor{rd2}{ We also compared the approach with a sparse recovery scheme that is known to reduce spatial variations in a reconstructed image, whilst preserving sharp discontinuities \cite{shaw2014}. We used a sparsity-promoting total-variation (TV) prior \cite{Rudin1992} implemented with Toast++ software \cite{schweiger2014toast++} and the conventional GN algorithm. The reconstructed images are shown in Figure \ref{fig:cross}(d). As seen, use of TV prior results in lower background variations compared to using Gaussian prior in Figure \ref{fig:cross}(b). However, the cross-talk artifact of the scattering inclusion in the absorption image is still visible. }

The study draws connections between statistical Bayesian and learning-based approaches, showing that insights from Bayesian inversion can be used in the design of learned image reconstruction. This opens the possibility for future studies to analyse these connections more carefully.

\section{Conclusions}\label{sec:conclusions}

We presented a novel approach for estimating absolute optical parameters in DOT, utilising a model-based iterative deep-learning approach. The results were validated with 2D simulations and a DOT experiment.
The results show that the proposed approach leads to improved computational times compared to conventional Gauss-Newton method. Also, the proposed approach can learn non-Gaussian image features and provide improved estimates for targets presenting sharp inclusions. No considerable loss of image quality was reported in situations where imaging targets did not match the training data. Furthermore, the proposed approach was shown to effectively compensate for modelling errors.
It was shown to provide improved computational time and estimation accuracy for non-Gaussian targets, compared to the Bayesian approximation error method. 

\section*{Acknowledgement}

The authors wish to acknowledge CSC – IT Center for Science, Finland, for computational resources.

\bibliographystyle{IEEEtran}
\bibliography{sample,newBib}

\end{document}